\documentclass[twocolumn,showpacs,preprintnumbers,amsmath,amssymb]{revtex4}

\usepackage{graphicx}
\usepackage{dcolumn}
\usepackage{bm}

\begin{document}

\title{Transition of multi-diffusive states in a biased periodic potential}

\author{Jia-Ming Zhang and Jing-Dong Bao\footnote{Corresponding author.
E-mail: jdbao@bnu.edu.cn}}

\affiliation{Department of Physics, Beijing Normal University,
Beijing 100875,  People's Republic of China}

\date{\today }

\begin{abstract}


We study a  frequency-dependent damping model of hyper-diffusion within the generalized Langevin equation.
The model allows for the colored noise defined by its spectral density, assumed to be proportional to $\omega^{\delta-1}$
at low frequencies with $0<\delta<1$ (sub-Ohmic damping) or $1<\delta<2$ (super-Ohmic damping), where the frequency-dependent damping is
deduced from the noise by means of the fluctuation-dissipation theorem.
It is shown that for super-Ohmic damping and certain parameters, the diffusive process of the particle
in a titled periodic potential undergos sequentially four time-regimes:
thermalization, hyper-diffusion, collapse and asymptotical restoration.  For analysing transition phenomenon of multi-diffusive states,
we demonstrate that the first exist time of the
particle escaping
from the locked state into the running state abides by an exponential distribution.
The concept of equivalent
velocity trap is introduced in the present model, moreover, reformation of ballistic diffusive system is also considered as a marginal situation,
however there does not exhibit the collapsed state of diffusion.

\end{abstract}

\maketitle

\section{Introduction}
\label{intro}

Diffusion in a periodic structure is of great interesting,
because it is simple to formulate and can be used to describe a surprising range of systems,
including Josephson junctions \cite{Barone1982}, charge-density waves \cite{Gruner1981}, superionic conductors \cite{Fulde1975}, rotation
of dipoles in an external field \cite{Reguera2000}, phase-locking loops \cite{Lindsey1972},
diffusion on surfaces \cite{Agassi1985}, and separation of particles by electrophoresis \cite{Ajdari1991,Nixon1996}.
The monumental work on this topic is to
be found in Risken's textbook \cite{Risken1984}, where the diffusion in a
washboard potential is analyzed within the framework of
the Fokker-Planck equation. Lately, with the term ¡°Brownian motion on a periodic substrate¡±, Marchesoni, H\"{a}nggi and collaborators have studied systemically
Gaussian noise-driven diffusion and transport in such potential \cite{hanrmd}. Understanding
particle diffusion in a one-dimensional system has been recognized as
a key issue in transport control, but then they did not refer necessarily to a point-like object diffusing
in a periodic potential \cite{tal}.  When pumping a dilute
mixture of interacting particles through a narrow channel \cite{bur},
either by applying external (dc or ac) gradients \cite{bor} or by
rectifying ambient fluctuations, the efficiency of the transport
mechanism is largely influenced by the diffusion of
the pumped particles.  In particular, the notions of hysteresis mechanism, multiple jumps, jump reversal, and backward-to-forward
rates were discussed in detail \cite{borprl1,borprl2}. Those models may explain various instances
of low-frequency excess damping in material science.

The forced Brownian
motion on periodic substrates  also provides an archetypal
model of transport in condensed phase. Analytical calculations and huge enhancement for effective diffusion coefficient
relative to the force-free case have been addressed
in the overdamped \cite{Reimann2001,Reimann2002} and underdamped
\cite{Marchenko2012,Marchenko2014,Sancho2010,Costantini1999,Lindner2016} cases.
In fact, the diffusion coefficient is  measured by the envelope width of the spatial distribution of the particle.
This is indeed realized through
the phenomenon of ``diffusion helped by transport".
The problem of whether or not one can change the scaling index of the coordinate variance increasing with time.
Recent studies on anomalous diffusion found across many different branches of physics
 are mostly to the absence of potential or the constant force case, the
 mean square displacement of the particle expresses
as $\langle x^2(t)\rangle\sim t^{\delta}$ with $0<\delta<2$.
An important aspect as regards understanding
a random process, in particular, an anomalous diffusive system, is its behavior in external fields.
To the best of our knowledge, however, unexpected
properties due to ``transport changing diffusive scaling law'' in a corrugated plane need to investigate in detail.

 Clearly, test particles in the unbiased periodic potential are distributed over many spatial wells at long times,
 so the diffusion changes eventually into normal one \cite{Risken1984,Siegle2010b}. Remarkably, our previous
 works showed that there exhibits two motion modes:
 the running state and the oscillating state for a super-Ohmic damping particle moving in a tilted periodic potential \cite{Bao2006,Lu2007}. The particle coordinate variance
 (CV) was approximately written as a power function of time, i.e.,
 $\langle \Delta x^2(t)\rangle\sim t^{\delta_{\mathrm{eff}}}$,
 where the index $\delta_{\textmd{eff}}$ can be enhanced up twice related to that of the force-free diffusive case.
 A minimal non-Markovian embedding model of ballistic diffusion was developed to interpret the hyper-diffusion phenomenon
and
 demonstrated that the transient hyper-diffusion will end up in the long-time limit \cite{Siegle2010a,sieb}.
Nevertheless,
 there exists a worry for transport that pervious numerical simulations may be not efficient to achieve
asymptotical results, moreover,
 for a more general non-Markovian dynamics subjected to no-Ohmic damping,
 the phenomenon of diffusion in a
washboard potential  remains unclear.

This paper is organized as follows. In Sec.~\ref{superdiffusion}, we evidence numerically the existence of four regimes for diffusion
of the super-Ohmic damping particle in a biased periodic potential. In Sec.~\ref{transformation}, we demonstrate that transformation from the locked state to the running
state obeys an
 exponential law and analyse the complicated diffusive behavior in the velocity space.
 Reformation of ballistic diffusion is discussed in Sec.~\ref{ballistic}. A summary is drawn in Sec.~\ref{summ}.

\section{The model and diffusive properties}
\label{superdiffusion}

We start out by considering a generalized Langevin equation (GLE) of a particle in a potential $V(x)$ and
subjected to
a memory damping and a thermal colored noise, i.e.,
\begin{equation}
\label{GLE}
m\ddot{x}+m\int_0^t\gamma(t-t')\dot{x}(t')dt'+V'(x)=\xi(t),
\end{equation}
where the noise of zero-mean obeys both the Gaussian distribution and the fluctuation-dissipation theorem $\langle\xi(t)\xi(t')\rangle=m k_B T \gamma(|t-t'|)$, $k_B$ is the Boltzmann constant, $T$ denotes the temperature, $\gamma(t)=\frac{2}{m\pi}\int_0^{\infty}d\omega \frac{J(\omega)}{\omega}\cos(\omega t)$, and $J(\omega)$  is
the bath spectral density corresponding to the system coupled bilinearly to a heat bath consisting of infinite harmonic oscillators
\cite{Zwanzig1973,Hanggi1990,Weiss1999,Kupferman2004}.
The GLE (\ref{GLE}) under a constant biased force $F$, thus, the potential is $V(x)=-Fx$, can be  solved  analytically \cite{Pottier2003,Morgado2002,Wang1999,Kupferman2004} for reference.
The mean displacement (MD) and the mean velocity (MV) of the particle are written as
\begin{equation}
\label{md}
\begin{aligned}
\langle x(t)\rangle=&\{x(0)\}+\frac{F}{m}\int_0^t H(\tau)d\tau,\\
\langle v(t)\rangle=&\{v(0)\}+\frac{F}{m}H(t).
\end{aligned}
\end{equation}
Herein we indicate by $\{\cdots\}$ the average with respect to the initial values of the state variables and by
$\langle\cdots\rangle$ we denote the average over the noise $\xi(t)$, respectively.
The CV and the velocity variance (VV) are given by
\begin{equation}
\label{msd}
\begin{aligned}
\langle\Delta x^2(t)\rangle=&\langle x^2(t)\rangle-\langle x(t)\rangle^2\\
=&\frac{2k_BT}{m}\int_0^t H(\tau)d\tau+\Big[\{ v^2(0)\}-\frac{k_BT}{m}\Big]H^2(t),\\
\langle\Delta v^2(t)\rangle=&\langle v^2(t)\rangle-\langle v(t)\rangle^2\\
=&\frac{k_BT}{m}+\Big[\{ v^2(0)\}-\frac{k_BT}{m}\Big]h^2(t),\\
\end{aligned}
\end{equation}
where $H(t)$ and $h(t)=\dot{H}(t)$ are two response functions and the Laplace transform of $H(t)$
is $\tilde{H}(s)=[s^2+s\tilde{\gamma}(s)]^{-1}$ \cite{Kubo1966}.

\begin{figure}[htbp]
\centering
  \includegraphics{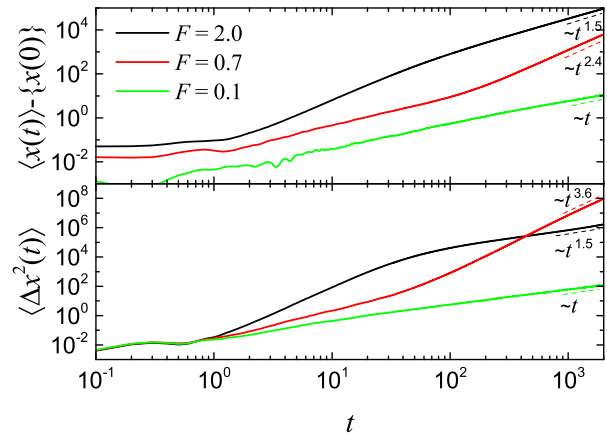}
\caption{(Color online) Time-dependent MD and CV of the particle for various biased forces
$F=2.0,~0.7,~0.1$
from top to bottom.}
\label{AD}
\end{figure}

Within the non-Ohmic damping (friction) model \cite{Weiss1999}, as an applied widely one of non-Markovian dynamics,  we have
   $J(\omega)=m\gamma_{\delta}\omega^{\delta}\exp(-\omega/\omega_c)$. Consequently, the index in the frequency-dependent damping appears to be
   linked, not only to $\delta$  as it would be the case in a heat bath, but also to characterize the color of noise.
   The Laplace transform of memory damping function $\gamma(t)$ is
  given by $\tilde{\gamma}(s)=\gamma_{\delta}/\sin(\delta\pi/2)s^{\delta-1}$ when $s\ll\omega_c$.
   The
low-frequency part of the spectral density governs long-time dynamics of the particle. Herein
the response function in Eqs. (2) and (3) is expressed as  $
H(t)\sim\sin(\delta\pi/2)(\gamma_{\delta}\Gamma(\delta))^{-1}t^{\delta-1}$  at long times \cite{Weiss1999}, and, thus
the asymptotical CV of the particle subjected to a constant force reads $\langle\Delta x^2(t)\rangle\sim 2D_{\delta}t^{\delta}/\Gamma(\delta+1)$,
where $D_{\delta}$ is the $\delta$-dependent diffusion constant.
In this work, we pay attention to diffusion of a particle in the following tilted  washboard potential:
\begin{eqnarray}
V(x)=-V_0\cos(2\pi x)-F x,
\end{eqnarray}
where  $V_0$ is the amplitude strength of periodic potential and $F$ is equal to constant. The potential $V(x)$ has
local minima if the driving force $F$ is smaller than the critical value, $F_c=2\pi V_0$, when $F>F_c$,
the minima vanish.

Unfortunately, the colored noise with non-Ohmic spectrum
 can not be
simulated directly, or, there is not a set of Markovian embedding equations
through introducing additional variables for GLE  (\ref{GLE}). In order to simulate transport process of a non-Ohmic damping particle
in a nonlinear potential, we develop an efficient method by using
the spectral approach  \cite{Lu2005,Romero1999,Makse1996,Banerjee2004} to generate the required colored noise
and the Runge-Kutta algorithm to solve numerically the GLE whole.
In the calculations, the natural units $m=1$ and $k_B=1$, the dimensionless parameters $V_0=1.0$ and $\gamma_{\delta}=1.0$;
the cutoff frequency  $\omega_c=10.0$, the time step $\Delta t=0.01$, as well as $5\times 10^3$ test particles are used. All the test particles start from a minimum of the titled washboard potential at $x(0)=(2\pi)^{-1}\arcsin[F/(2\pi V_0)]$ and their velocities are sampled from the
Gaussian distribution with zero-mean and width $\{ v^2(0)\}=k_BT/m$.

In Fig. 1, we reveal both super-current and super-diffusion of the particle for $\delta=1.5$.
For a large titled force $F=2.0$ but being less than the critical value $F_c$, during a long-time window, super-diffusion
appears, however, for a very small force $F=0.1$,
there exhibits normal diffusion.
For a moderate force $F=0.7$, after a
short initial period of time processing normal diffusion, both MD and CV increase with time faster than $t^{1.5}$ as $t^{\delta_{\mathrm{eff}}}$ with
effective power indices $\delta_{\mathrm{eff}}=2.4$ and $\delta_{\mathrm{eff}}=3.6$, respectively. Nevertheless, the hyper-diffusion of this kind \cite{Lu2007,Siegle2010b,Siegle2010a} is only a transient behavior.
 If the tilted force is very large, for example, $F=2.0$, the time-regime of hyper-diffusion  extends approximately over one decade from $t=1$ to $t=10$ with $\delta_{\mathrm{eff}}>1.5$, but then turns into the  super-diffusion of $\delta_{\mathrm{eff}}=1.5$.

\begin{figure*}[htbp]
\centering
  \includegraphics{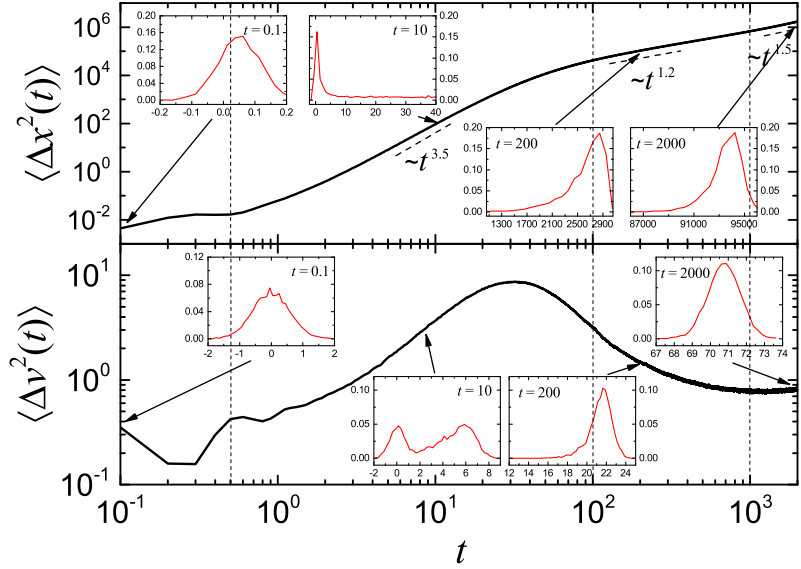}
\caption{Time-dependent CV and VV of the particle.
Four regimes are divided with dashed vertical lines: $\langle\Delta x^2(t)\rangle\sim t^{\delta_{\mathrm{eff}}}$, the thermalized regime $\delta_{\mathrm{eff}}\sim 0$, the hyper-diffusive regime $\delta_{\mathrm{eff}}>\delta=1.5$, the collapsed regime $\delta_{\mathrm{eff}}<\delta$, as well as the asymptotical regime $\delta_{\mathrm{eff}}=\delta$. The
parameters used are $F=2.0$, $T=0.5$ and $\delta=1.5$. }
\label{X2}
\end{figure*}

In Fig.~\ref{X2}, we plot time-dependent CV and VV of the particle.
It is seen that the diffusive process can be distinguished into four time regimes:
thermalization, hyper-diffusion, collapse and asymptotical restoration.
 Snapshots of the particle coordinate and velocity distributions at $t=0.1, 10, 200, 2000$ are presented for various
regimes. In the thermalized regime, both CV and VV have small changes related to their initial values, and thus, the distributions of coordinate and velocity of the particle are
always Gaussian forms.
 Then diffusive process evolves into the hyper-diffusive regime. The particle gains enough energy to overcome dissipation
 loss and thus moves along the direction of biased force. This results in a long-tail in front the peak of coordinate distribution
 at time $t=10$.

 The above phenomena can be understood well in the velocity space. Up inspection, we find a prominent
 result: The VV of the particle becomes rapidly large and the velocity
 distribution exhibits a bimodal structure. Namely, with time increasing, the test particles give  gradually rise from the initial
 Gaussian velocity distribution, which is regarded as the locked state. For the running state, consequently,  the peak of the coordinate distribution
  moves with a certain acceleration. With test
  particles escape continuously  from the locked state into the running state, the VV
  arrives gradually at its maximum and then begins on declining. The CV increases with time slower than before, but follows an effective
   diffusive index $\delta_{\mathrm{eff}}$, which is still larger than $1.5$. This implies that the hyper-diffusive regime may last a long time for smaller biased forces until numerical simulation finished.

 The criteria for the end of the second regime is that all the test particles escape from the initial locked state to the running state. Next, the third regime, i.e., the collapsed one rises to appear. All the test particles move toward to the direction of biased force and none is trapped by the potential well again. The test particles with lower velocities will get the energies larger than that of the faster ones, so that
the CV of the particle collapses with an effective index $\delta_{\mathrm{eff}}=1.2$ in comparison with
$\delta=1.5$. In the velocity space,
the tail behind the peak of velocity distribution
shortens gradually with the increase of time. Finally, the asymptotical regime comes in turn. The velocity distribution of the particle returns to a Gaussian form, which gives
rise to the diffusion returning to the super-diffusive state with $\delta_{\mathrm{eff}}=1.5$.

\section{Transition of velocity from locked state to running state}
\label{transformation}

As demonstrated herein, due to double modes of the velocity distribution, we observe the strong amplification of diffusive behavior in the underdamped case \cite{Lindner2016,Costantini1999,Lindenberg2005}. We also find  from Fig.~\ref{X2} that both
the locked state and the running state of the particle velocity exist and expect to interpret the complicated phenomena of hyper-diffusion. It is worth noticed that for the super-Ohmic damping
particle, its average velocity is about equal to zero in the locked state, and the running state is ``running" with a velocity increasing with time
 rather than a constant value.
 We admit that the velocity of the particle in the running state emerges as $v(t)\approx F\sin(\delta\pi/2)t^{\delta-1}/[m\gamma_{\delta}\Gamma(\delta)]$ according to Eq.~(\ref{md}). Once test particles enter the running state, the average kinetic energy of the particle gained from the
 external driving is larger than the dissipated energy due to the memory damping effect,
 so these test particles will never return to the locked state. A burning question is that how the
 particle transforms from the locked state to the running state?

\begin{figure}[htbp]
\centering
  \includegraphics{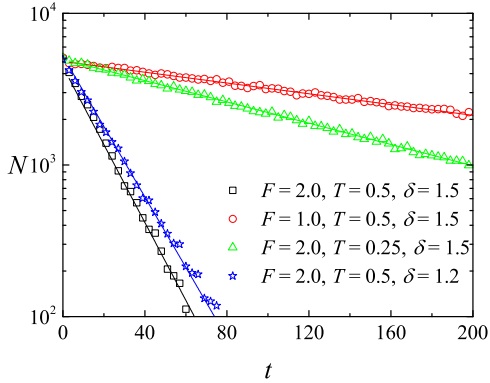}
\caption{(Color online) Log-lin plotting of number of test particles remaining in the locked state for various values of
external force, temperature and power index.}
\label{NS}
\end{figure}

 In Fig.~\ref{NS},
we count negative velocity of $5\times 10^3$ test particles and take twice the value as the number of test
particles remaining in the locked state
for various values of $T$, $F$ and $\delta$.
The exponential characteristic of proportion decay in the locked state is vary alike the escape process from a metastable potential,
 thus we put forward a concept of equivalent velocity ``trap''.  The locked state is considered as the one where test particles are confined in the velocity trap, and the
running state is recognized as the one where test particles are driven by an
 effective force $f(t)\approx F\sin(\pi \alpha/2)t^{\alpha-2}/[m\gamma_{\alpha}\Gamma(\alpha)]$. This is
 similar to the heuristic treatment of Ref.~\cite{Marchenko2014}.
  In the hyper-diffusive time-regime, the
 escape of the particle from a spatial well can represent the effect of multiple scattering on
  the barriers of periodic potential.

Up observing Fig. 3,
we assume that the first exit time $t_e$ of the particle escaping from the locked state follows an exponential distribution, i.e.,
\begin{equation}
p(t_e)=\frac{1}{t_0} \exp\left(-\frac{t_e}{t_0}\right).
\end{equation}
The mean exit time $t_0$ can be extracted from the numerical results.  It is seen from Fig.~\ref{NS} that $t_0=16.3$ under
 the parameters: $F=2.0, T=0.5, \delta=1.5$; $t_0=244.0$ under $F=1.0,~ T=0.5,~ \delta=1.5$; $t_0=124.1$ under
$F=1.0,~ T=0.5,~ \delta=1.5$; and $t_0=19.4$ under  $F=1.0,~ T=0.5,~ \delta=1.5$. At time $t$,
the distribution of running time $t_r$ of the particle after escaping over the locked state
is defined by $p_r(t_r)=t_0^{-1}\exp[-(t-t_r)/t_0]$ with $t_e+t_r=t$.

 The temporal distribution of the first exit time can be transformed into a normalized velocity distribution
 via $v=F\sin(\delta\pi/2)t_r^{\delta-1}/[m\gamma_{\delta}\Gamma(\delta)]$, we obtain
\begin{equation}
\label{pvr1}
\begin{split}
p_r(v)=&\frac{1}{\delta-1}v_0^{-\frac{1}{\delta-1}}v^{\frac{2-\delta}{\delta-1}}\\
&\times\frac{\exp\Big[(v/v_0)^{1/(\delta-1)}-(v_t/v_0)^{1/(\delta-1)}\Big]}
{1-\exp\Big[-(v_t/v_0)^{1/(\delta-1)}\Big]},
\end{split}
\end{equation}
where $v_0=F\sin(\delta\pi/2)t_0^{\delta-1}/[m\gamma_{\delta}\Gamma(\delta)]$ and $v_t=F\sin(\delta\pi/2)t^{\delta-1}/[m\gamma_{\delta}\Gamma(\delta)]$.
 In fact, there are two contributions to the width of $p_r(v)$. One is the expositional
 distribution due to dispersion of emergence time from the velocity trap,
 and the other comes from the thermal fluctuation. Considering both of them, the modified velocity distribution
of the particle in the running state is written as
\begin{equation}
\label{pvr2}
p_r^m(v)=\int_0^{v_t}\frac{1}{\sqrt{2\pi}\sigma}\exp\Big[-\frac{(v-v')^2}{2\sigma^2}\Big]p_r(v')dv',
\end{equation}
where $\sigma^2=k_BT/m$.

In the spirit of relation between the group diffusion and the phase diffusion \cite{Bao2013},
 the total probability density function (PDF) $p(v)$ is composed of two sub-PDFs: $p_l(v)$ (the locked state)
  and $p_r(v)$ (the running state).
 We yield $p(v)=a(t) p_l(v)+[1-a(t)] p_r(v)$, where $a(t)=\exp(-t/t_0)$ is a proportion of test particles in the locked state.
  Then, the total MV and VV are given by $\langle v\rangle=a(t) \langle v\rangle_l+[1-a(t)]\langle v\rangle_r$ and $\langle\Delta v^2\rangle=a(t)\langle\Delta v^2\rangle_l+[1-a(t)]\langle\Delta v^2\rangle_r+a(t)[1-a(t)][\langle v\rangle_r-\langle v\rangle_l]^2$. Here, $\Delta v=v-\langle v\rangle$, $\langle\cdots\rangle_l$ and $\langle\cdots\rangle_r$ denote statistical averages of test
  particles in the locked and running states, respectively. The particle velocity distribution in the locked state is a Gaussian function,
  and its stationary form reads $p_l(v)=\exp[-v^2/(2\sigma_l^2)]/(\sqrt{2\pi}\sigma_l)$. It has zero-mean and variance $\langle\Delta v^2\rangle_l=\sigma_l^2\approx k_BT/m$. In the running state, the particle velocity distribution $p_r(v)$ has been determined by Eq.~(\ref{pvr1}) for simplicity, where the
  MV is given by
\begin{equation}
\begin{split}
\langle v\rangle_r=&\int_0^{v_t} vp(v)dv\\
=&v_0 \exp(-i\pi\delta)\gamma\Big[\delta,-(v_t/v_0)^{1/(\delta-1)}\Big]\\ &\times\frac{\exp\Big[-(v_t/v_0)^{1/(\delta-1)}\Big]}{1-\exp\Big[-(v_t/v_0)^{1/(\delta-1)}\Big]},
\end{split}
\end{equation}
and the mean square velocity emerges as
\begin{equation}
\begin{split}
\langle v^2\rangle_r=&\int_0^{v_t} vp_r(v)dv\\
=&-v_0^2\exp(-2i\pi\delta)\gamma\Big[2\delta-1,-(v_t/v_0)^{1/(\delta-1)}\Big]\\ &\times\frac{\exp\Big[-(v_t/v_0)^{1/(\delta-1)}\Big]}{1-\exp\Big[-(v_t/v_0)^{1/(\delta-1)}\Big]},
\end{split}
\end{equation}
where $\gamma[\cdot,\cdot]$ is the incomplete gamma function. Noticed that $\langle\Delta v^2\rangle_r=k_BT/m+\langle v^2\rangle_r-\langle v\rangle_r^2$ , which comes simply from the sum of thermal fluctuation and dispersion effect.
In this way, we obtain the total VV, reading
\begin{equation}
\label{v2}
\begin{split}
\langle\Delta v^2(t)\rangle=&\frac{k_B T}{m}-\frac{F^2 t_0^{2\delta-2}\sin{\frac{\pi \delta}{2}}}{m^2\gamma_{\delta}^2\Gamma^2(\delta)}e^{-2i\pi\delta}\\ &\times\Big[\gamma^2(\delta,-t/t_0)e^{-\frac{2t}{t_0}}+\gamma(2\delta-1,-t/t_0)e^{-\frac{t}{t_0}}\Big],
\end{split}
\end{equation}
which will be used to analyse the hyper-diffusion.

\begin{figure}[htbp]
\centering
  \includegraphics{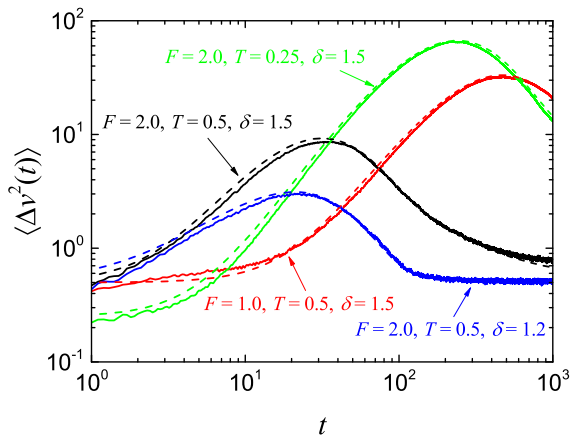}
\caption{(Color online) The VV of $5\times 10^3$ test particles simulated (solid line) and theoretical (dashed line)
results for comparison. Theoretical mean exit time $t_0$ is extracted from Fig.~\ref{NS} and the parameters used
are the same as Fig.~\ref{NS}. }
\label{NV2}
\end{figure}

In Fig.~\ref{NV2}, we depict the theoretical VV for various parameters and compare
with numerical results. It is seen that our analytical results are in well agreement with the simulations. At beginning,
the total VV approaches $\sigma_l^2$. When the test particles enter the hyper-diffusive regime, Eq.~(\ref{v2}) can be locally expanded as $\langle\Delta v^2(t)\rangle\propto t^{\lambda}$. This result evidences that the CV of the particle
emerges as
 $\langle\Delta x^2(t)\rangle\propto t^{2+\lambda}$, the hyper-diffusion appears. At enough long times, all the test particles enter into the running state, the particle
velocity distribution due to dispersion will gradually collapse to a $\delta$-peaked distribution. Finally, the VV
 returns to the thermal equilibrium value $\lim_{t\to\infty}\langle\Delta v^2(t)\rangle\approx k_BT/m$.
 Based on the above facts, we conclude that although super-diffusion has a complicated intermediate transient process, the effect of dispersion by the periodic potential vanishes at long times.

\section{Reformation of ballistic diffusive system}
\label{ballistic}

The ballistic diffusion of a force-free particle is indeed the limitation of thermal diffusion described by GLE.
Differing from super-diffusion in terms of the non-Ohmic damping model with $1<\delta<2$,  the memory to the initial velocity preparation
does not vanish and consequently the velocity variable is a nonergodic one
because of vanishing Markovian low-frequency friction \cite{Siegle2010b}, which is regarded as a marginal case of our model.
A simple but physically reasonable model for the memory damping function is
chosen to be
$\gamma(t)=\gamma[2\delta(t)-\nu e^{-\nu t}]$ \cite{Bao2005,Siegle2010a}, which corresponds to the
spectral density of heat bath given by $J(\omega)$ is cubic when $\omega\ll\nu$. The non-Markovian GLE of this kind can
allow for a three-dimensional Markovian LE embedding \cite{Bao2005,Siegle2010a}:
\begin{equation}
\begin{aligned}
\dot{x}(t)=&v(t),\\
m\dot{v}(t)=&-V'(x)-u(t)-m\gamma v(t)+\sqrt{2mk_B T\gamma}\zeta(t),\\
\dot{u}(t)=&-m\nu\gamma v(t)-\nu u(t)+\nu\sqrt{2mk_B T\gamma}\zeta(t),
\end{aligned}
\end{equation}
where $\zeta(t)$ is a zero-mean Gaussian white noise and $\langle\zeta(t)\zeta(t')\rangle=\delta(t-t')$. The initial value
of the auxiliary variable $u(0)$ obeys the Gaussian distribution with zero-mean and variance $\langle u^2(0)\rangle=mk_BT\nu\gamma$.

Similar to analysis in the above section, we yield all the test particles with the velocity
$v\approx F\nu t/[m(\gamma+\nu)]$ at long times after they escape from the initial
 velocity trap.
The distribution of the first exit time from the well is still: $p(t_e)=t^{-1}_0 \exp(-t_e/t_0)$ and it can be transformed into the
 normalized velocity distribution: $p_r(v)=v_0^{-1}\exp[(v-v_t)/v_0]/[1-\exp(-v_t/v_0)]$ ($0\leq v\leq v_t$) for the test particles in the running state. The
particle velocity distribution in the locked state still remains a Gaussian form, we get the total VV given by
\begin{equation}
\label{bv2}
\begin{split}
\langle\Delta v^2(t)\rangle=&\frac{k_B T}{m}+\frac{F^2\nu^2 t_0^2}{m^2(\gamma+\nu)^2}\\
&\times\left[1-\exp\left(-\frac{t}{t_0}\right)-2\exp\left(-\frac{t}{t_0}\right)\frac{t}{t_0}\right].
\end{split}
\end{equation}

In Fig. 5(a), we compare the simulation result with the theoretical one with the parameters used in Ref.~\cite{Siegle2010a}, i.e.,
 $\gamma=1.0$, $\nu=0.25$, $V_0=1.0$ and $\Delta t=10^{-4}$.
Likewise, the mean exit time $t_0$ is also extracted by numerical statistics. The results are listed below: $t_0=615.13$ ($F=0.75, T=0.5$), $t_0=33.48$ ($F=1.5, T=0.5$) and $t_0=348.64$ ($F=1.5, T=0.25$).  The mean square velocity rises gradually  to the
stationary value and no collapsed process exists in comparison with the super-diffusion for $1<\delta<2$.

\begin{figure}[htbp]
\centering
  \includegraphics{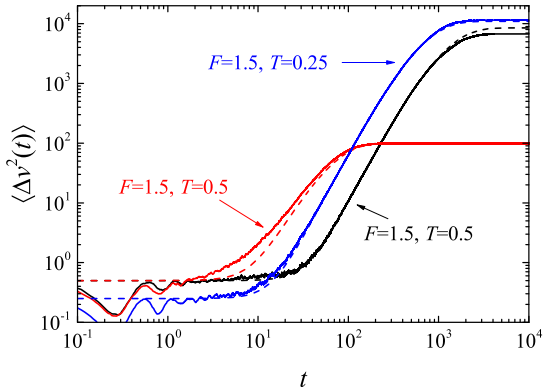}
\caption{(Color online) The VV of $10^4$ test particles simulated (solid line) \cite{Siegle2010a} and theoretical Eq.~(\ref{bv2})
(dashed line) for different values of $F$ and $T$. }
\label{BV}
\end{figure}

The ``kinetic temperature" notion $T_{\textmd{kin}}$ was used to characterize the width of
a nonequilibrium velocity distribution  \cite{Siegle2010a}. Here, the average kinetic energy of the particle can be decomposed as $K=K_m+K_T$, where $K_m$ represents the kinetic energy of the particle under a constant force and $K_T$ characterizes multiple scattering on the barriers of periodic potential. Our theoretical result is associated with the effective kinetic temperature by $K_m=k_BT/2$ and $K_T= k_BT_{\textmd{kin}}/2=F^2\nu^2 t_0^2[1-\exp(-t/t_0)-2t\exp(-t/t_0)/t_0]/[2m(\gamma+\nu)^2]$. If all the particles escape from the initial velocity trap, they arrive at the maximal kinetic temperature $T_{\mathrm{max}}$. In our analyses,
 the mean exit time $t_0$ of the particle from the velocity trap determines the crossover time from the locked state to the
 running state and the VV depends mainly on it. It is an important parameter and depends strongly  more than exponential form on both the temperature and the biased force, similar to the kinetic temperature.
The long-time limitation of Eq.~(\ref{bv2}) yields $\lim_{t\to\infty}\langle\Delta v^2(t)\rangle\approx k_BT/m+F^2\nu^2 t_0^2/[m(\gamma+\nu)]^2$. This result is composed by thermal fluctuation and multiple scattering on the barriers of periodic potential, it
 differs markedly with that of the usual super-diffusion, the latter contains only the thermal fluctuation.
 This implies that the dispersion effect will not collapse  for a ballistic diffusion system, so there is no the
 collapsed regime. Therefore, the CV of the particle will turn to that of
 the ballistic diffusion from hyper-diffusion situation.

\section{Summary}
\label{summ}

We have researched within the GLE-formalism
diffusive dynamics of a non-Ohmic damping particle in a titled periodic potential.
The emerging non-equilibrium features are manifested by the model parameters and process-dependent diffusive scaling law.
It has
found that the diffusion of the super-Ohmic damping particle might undergo four time regimes:
thermalization, hyper-diffusion, collapse and asymptotical restoration.
In order to understand rich properties of hyper-diffusion, we propose a velocity trap
 that the motion is composed of two parts: (i) the locked state, where the particle in captured in a potential well;
 (ii) the running state, where the particle gains enough energy to escape the potential well. In particular,
 the latter seems as if the particle is
 driven by a time-dependent force.
 We have demonstrated that the escaping time of the particle from the locked state follows an exponential distribution law.
 This leads to the dispersion behavior and induces the hyper-diffusion.
 Moreover, the ballistic diffusion found in previous works is also considered as a marginal case of our model, where the velocity variance of the
 particle contains only the thermal fluctuation and no dispersion. It has
shown that the dispersion effect will not collapse, so there is no the collapsed regime.

The present study may be related to self-propelled motion or active
Brownian motion. This is one of the key features of life appearing on levels ranging from flocks of animals to single
cell motility and intracellular transport by molecular motors \cite{Lindner2008,Berg1993,Bray2001,Howard2001}.
We are also confident that our results for the biased periodic potential induced hyper-diffusion
will serviceably impact other properties of the
washboard-potential device and quantum diffusion. Thus
 this field is open for future that in turn reveal surprising findings.

\section*{ACKNOWLEDGMENTS}

This work was supported by  the
National Natural Science Foundation of China under Grant No. 11575024.

\end{document}